\documentclass[pra,aps,twocolumn,showpacs]{revtex4}
\usepackage[OT4]{fontenc}

\usepackage{amsmath,amsfonts}
\usepackage{graphicx}

\usepackage{bm}

\newcommand{\rr}{\bm{r}}

\newcommand{\psic}{\psi_{\mathrm{c}}(\bm{r})}
\newcommand{\Psiv}{\Psi_{\mathrm{v}}(\bm{r})}
\newcommand{\Psic}{\Psi_{\mathrm{c}}(\bm{r})}
\newcommand{\lon}{\mathrm{l}}
\newcommand{\tra}{\mathrm{t}}

\DeclareMathOperator{\im}{Im}

\begin{document}
\author{P. Kaczmarkiewicz}
\author{A. Musia{\l}}
\author{G. S\k{e}k}
\author{P. Podemski}
\author{P. Machnikowski}
\author{J. Misiewicz}
\affiliation{Institute of Physics, Wroc{\l}aw University of
Technology, 50-370 Wroc{\l}aw, Poland}

\title{Hole subband mixing and polarization of luminescence from quantum
  dashes: a simple model}

\begin{abstract}
In this paper, we address the problem of luminescence polarization in
the case of nanostructures characterized by an in-plane shape
asymmetry. We develop a simple semi-qualitative model revealing the
mechanism that accounts for the selective polarization properties of
such structures. It shows that they are not a straightforward
consequence of the geometry but are related to it via valence
subband mixing. Our model allows us to predict the degree of
polarization (DOP) dependence on the in-plane dimensions of investigated
structures assuming a predominantly heavy hole character of the valence band
states, simplifying the shape of confining potential and neglecting
the influence of the out-of plane dimension. The energy dependence
modeling reveals the importance of different excited states in
subsequent spectral ranges leading to non-monotonic
character of the DOP. The modeling results show
good agreement with the experimental 
data for an ensemble of InAs/InP quantum dashes for a
set of realistc parameters with the heavy-light hole states separation being
the only adjustable one. All characteristic features are reproduced in
the framework of the proposed model and their origin can be well explained
and understood. We also make some further predictions about the
influence of both the internal characteristics of the nanostructures
(e.g. height) and the external conditions (excitation power, temperature)
on the overall DOP.   
\end{abstract}


\maketitle

Quantum dashes (QDashes)
\cite{utzmeier96,sauerwald05,loffler06,rudno06,hein08,sek09}
are quai-zero-dimensional nanostructures characterized by
strong in-plane asymmetry. Typically, their width is on the order of
several to a few tens of nanometers, while their length may be on the
order of hundreds of nanometers
\cite{utzmeier96,sauerwald05,loffler06,rudno06}. 
The electronic structure and optical properties of these structures
have been investigated both 
experimentally \cite{dery04,rudno06,hein08,sek09,jo10}
and theoretically \cite{wei05,planelles09,miska04,dery04}, which is
motivated by their favorable properties from the point of view of
photonic applications at telecom wave lengths \cite{hein08}. For those
and other applications, the understanding of polarization properties
(expected to be anisotropic due to shape asymmetry) is crucial.

\begin{figure}[tb]
\begin{center}
\includegraphics[width=85mm]{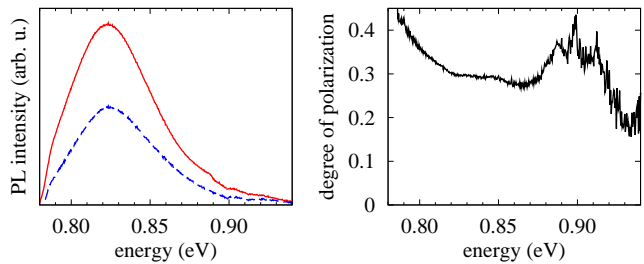}
\end{center}
\caption{\label{fig:exp}(a) Room temperature photoluminescence spectra
  of an ensemble of 
  QDashes at two orthogonal linear  polarizations (red solid line:
  parallel to the structure elongation, blue dashed line: perpendicular to
  the structure elongation). (b) The corresponding degree
  of polarization. The apparent structure in the high-energy part
  results from irrelevant fluctuations
  due to water vapor absorption in the aera of weak photoluminescence signal.}
\end{figure}
 
In Fig.~\ref{fig:exp}(a) we show the polarization-dependent
photoluminescence spectrum obtained at $300$~K from an 
ensemble of epitaxially grown self-assembled InAs/InP QDashes
preferentially elongated in the 
$[1\overline{1}0]$ direction,
similar to those studied in Ref.~\cite{sauerwald05,rudno06}. 
Fig.~\ref{fig:exp}(b) shows the corresponding degree of polarization (DOP),
\begin{displaymath}
P=\frac{I_{||}-I_{\bot}}{I_{||}+I_{\bot}},
\end{displaymath}
where $I_{||},I_{\bot}$ are the intensities of luminescence components
polarized along and perpendicular to the QDash elongation axis. 
The anisotropic shape of the QDashes leads to a high
DOP of the emitted radiation (see also Ref.~\cite{jo10}).
This polarization cannot
be a trivial consequence of the 
confinement shape at the level of a single-band effective mass and
envelope function description 
as the emission is governed by the interband dipole
moment that depends on the Bloch parts of the wave functions and not
on their envelopes. Therefore, a more general description is
needed in order to take subband mixing into account. 
Quantitatively exact modeling of such non-trivial optical properties
of
nanostructures must be based on atomistic or multi-band kp methods
\cite{planelles09,andrzejewski10}. However, it may be useful and
interesting to have a simple semi-quantiative model that would
not only elucidate the physical mechanism of the polarization but also
yield an estimate of the expected effect in terms of the shape
parameters. 

In this contribution, we present a ``minimal'' theory that is able to
account for the observed DOP in the QDash luminescence.  
First, we show that the polarization is
proportional to the degree of heavy-light hole mixing. Then we estimate
the degree of the latter within a semi-quantitative model and show
that it is related to the shape asymmetry of the structure. We provide
a very simple formula relating the DOP to the QDash
dimensions. 

The polarization of light emitted in a recombination process from a
conduction band (cb) state $\Psic$ to a valence band (vb) state $\Psiv$ is
determined by the interband matrix element of the dipole moment operator
$\hat{\bm{d}}$,
\begin{displaymath}
\bm{d}=\int d^{3}r\Psi_{\mathrm{v}}^{*}(\rr)\hat{\bm{d}}\Psic
=\sum_{\lambda}a_{\lambda}\bm{d}_{\lambda}.
\end{displaymath}
Here we have performed the standard separation of length scales in
the integration, $\bm{d}_{\lambda}$ is the
bulk interband dipole moment between the initial cb (say, with spin
up) and the final valence subband, and
\begin{displaymath}
a_{\lambda}=\int 
d^{3}r\psi_{\lambda}^{*}(\rr)\psic
\end{displaymath}
is the envelope function overlap,
where we denoted the cb envelope wave function by $\psic$
and the components of the vb envelope by $\psi_{\lambda}(\rr)$, where
$\lambda$ is the subband index.

For the initial state in the spin-up conduction subband, only two
transitions yield non-vanishing in-plane components of the dipole
moment: to the $+3/2$ heavy hole band and to the $-1/2$ light hole
band. The corresponding bulk matrix elements are
\begin{displaymath}
\bm{d}_{3/2}=
\frac{d_{0}}{\sqrt{2}}\left(  \begin{array}{r}
 -1 \\i    \end{array} \right),\quad
\bm{d}_{-1/2} 
= 
\frac{d_{0}}{\sqrt{6}}\left(  \begin{array}{r}
1 \\ i 
  \end{array} \right).
\end{displaymath}
where the two components of the in-plane vectors refer to the
Cartesian coordinates $x,y$ bound to the crystallographic axes.

In order to study the properties of a nanostructure elongated along
the $[1\overline{1}0]$ direction we define the unit vectors along
and transverse to the elongation direction
\begin{displaymath}
\hat{\bm{e}}_{\lon}=\frac{\hat{\bm{e}}_{x}-\hat{\bm{e}}_{y}}{\sqrt{2}},
\quad
\hat{\bm{e}}_{\tra}=\frac{\hat{\bm{e}}_{x}+\hat{\bm{e}}_{y}}{\sqrt{2}}.
\end{displaymath}
The components of the interband dipole moment along these directions
are 
\begin{equation}\label{dl}
d_{\lon}=-d_{0}\frac{i+1}{2}a_{3/2}
+d_{0}\frac{1-i}{2\sqrt{3}}a_{-1/2}
\end{equation}
and
\begin{equation}\label{dt}
d_{\tra}=d_{0}\frac{i-1}{2}a_{3/2}
+d_{0}\frac{i+1}{2\sqrt{3}}a_{-1/2}.
\end{equation}
Note that we use the standard definition of the basis
functions \cite{sakurai94,winkler03} (which differs from that used in
many papers employing the kp theory
\cite{pryor98,andrzejewski10}). 

The intensities of the linearly polarized components of the emitted
radiation along and transverse to the structure are proportional to
$|d_{\lon}|^{2}$ and $|d_{\tra}|^{2}$, respectively.
Hence, the DOP is 
$P=(|d_{\lon}|^{2}-|d_{\tra}|^{2})/(|d_{\lon}|^{2}+|d_{\tra}|^{2})$. From
Eqs.~\eqref{dl} and \eqref{dt}, one finds
\begin{displaymath}
\left|d_{\lon}\right|^{2}+\left| d_{\tra}\right|^{2}=
\left|d_{0}\right|^{2}\left[ 
\left|a_{3/2}\right|^{2}+\frac{1}{3}\left|a_{-1/2}\right|^{2}
 \right] 
\end{displaymath}
and
\begin{displaymath}
\left|d_{\lon}\right|^{2}-\left| d_{\tra}\right|^{2}=
-\frac{2}{\sqrt{3}}\left|d_{0}\right|^{2} \im\left[ 
a_{3/2}^{*} a_{-1/2} \right].
\end{displaymath}
From these equations, it is clear that mixing between heavy and light
hole contributions in the confined hole state can lead to anisotropy of
emission polarization with respect to the structure geometry, depending on the
relative phase of the light-hole and heavy-hole components.

A more quantitative conclusion may be achieved if one assumes that the
lowest hole state is predominantly of heavy hole character, with an
admixture of light hole states. This is justified in many structures
since the light hole states are shifted in energy with respect to the
heavy hole states due to confinement and strain and the interband
coupling elements are relatively 
small. It follows from the structure of the 
Kane hamiltonian \cite{pryor98,winkler03} that the heavy hole state
with the angular momentum $+3/2$ is coupled in the leading order of
perturbation to both light hole states. However, the coupling term
between this heavy hole state and the $+1/2$ state is proportional to
$k_{z}$ which means that only states excited along the $z$ direction
are coupled. Due to the strong confinement in the growth direction these
states have a very high energy and their contribution is expected to
be small. Thus, the hole state is essentially composed of the $+3/2$
hh component with an admixture of a $-1/2$ lh component. In the 1st
order of perturbation theory one finds for the lh admixture
\begin{eqnarray*}
\psi_{-1/2}(\rr) & = & \sum_{n}
\frac{1}{\Delta E_{\mathrm{lh}}}
d^{3}r' \phi_{n}^{*}(\rr')V\psi_{3/2}(\rr')\phi_{n}(\rr)\\
 & = & \frac{1}{\Delta E_{\mathrm{lh}}}V \psi_{3/2}(\rr),
\end{eqnarray*}
where 
$\{\phi_{n}\}$ is any complete set of functions, 
$V$
is the relevant inter-subband element of the kp Hamiltonian, 
$\Delta E_{\mathrm{lh}}=E_{\mathrm{h}}-E_{\mathrm{l}}$ is the energy
separation between the heavy and light hole subbands,
the differences between energies of various lh states have been
neglected, and the last equality is obtained from the completeness
relation 
\begin{displaymath}
\sum_{n} \phi_{n}^{*}(\rr')\phi_{n}(\rr)=\delta(\rr-\rr').
\end{displaymath}

With our choice of basis states, the element of the Kane Hamiltonian
coupling the relevant states is \cite{winkler03}
\begin{displaymath}
V=\frac{\sqrt{3}\hbar^{2}}{2m_{0}}\left[
  \gamma_{2}(k_{x}^{2}-k_{y}^{2})
+2i\gamma_{3}k_{x}k_{y} \right],
\end{displaymath}
where $k_{j}=i\partial/\partial x_{j}$, $m_{0}$ is the free electron
mass and $\gamma_{j}$ are Luttinger
parameters. Only the imaginary part of $V$ contributes to 
$\im\left[a_{3/2}^{*} a_{-1/2} \right]$,
\begin{displaymath}
\im V=-\frac{\sqrt{3}\hbar^{2}}{2m_{0}}\gamma_{3}
(k_{\mathrm{l}}^{2}-k_{\mathrm{t}}^{2}),
\end{displaymath}
where we used the components relative to the structure elongation.

In order to obtain a general estimate without relying on specific
information on
the confined wave functions we note that the dominating contribution
to luminescence originates from electron and hole wave functions with
the same quantum numbers and neglect the differences between their
exact shape, so that $a_{3/2}\approx 1$. Assuming the simplest
rectangular box model for the confinement we find
\begin{eqnarray*}
\lefteqn{\im a_{-1/2}=}\\
&&\int d^{3}r \psic V^{*}\psi_{3/2}^{*}(\rr)\approx
-\frac{\sqrt{3}\hbar^{2}\pi^{2}}{2m_{0}}\gamma_{3}
\left( \frac{n_{\tra}^{2}}{D^{2}}-\frac{n_{\lon}^{2}}{L^{2}} \right),
\end{eqnarray*}
where $L,D$ are the dash length and width, respectively and 
$n_{\lon},n_{\tra}$ are the corresponding quantum numbers in the
rectangular box model. Hence, the DOP is
\begin{equation}\label{pol}
P=\frac{\hbar^{2}\pi^{2}}{m_{0}}\gamma_{3}
\left( \frac{n_{\tra}^{2}}{D^{2}}-\frac{n_{\lon}^{2}}{L^{2}} \right),
\end{equation}

\begin{figure}[tb]
\begin{center}
\includegraphics[width=85mm]{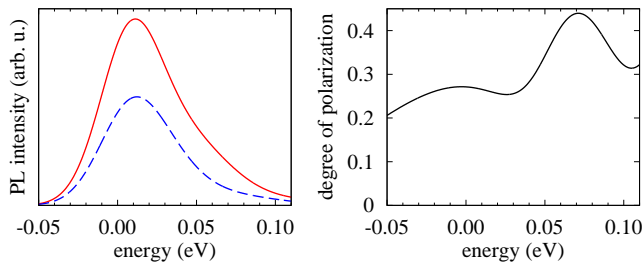}
\end{center}
\caption{\label{fig:theor}(a) Theoretical results for the room
  temperature photoluminescence 
  spectra of an ensemble of 
  QDashes at two orthogonal linear polarizations (red solid line:
  parallel to the structure elongation, blue dashed line: perpendicular to
  the structure elongation). (b) The corresponding degree
  of polarization. The energies are shown relative to the ground state
  at the maximum of the distribution.}
\end{figure}

The formula \eqref{pol} is the main theoretical result of our
study. In spite of its simplicity and approximate character, it
reproduces experimental data
reasonably, as can be seen by comparing 
the measured DOP shown in Fig.~\ref{fig:exp} with
the results of theoretical modeling presented in Fig.~\ref{fig:theor} . Here,
we plot the results obtained from an ensemble of QDashes modeled by a
set of boxes (with infinite potential walls) characterized
by the fixed height to width ratio of $H/D=1/3$, identical length
$L=150$~nm and a gaussian distribution of ground state energies with
the standard deviation of $44$~meV (which uniquely determines the size
distribution). This set of parameters corresponds to the geometry of
the QDash ensemble \cite{sauerwald05,rudno06} and to the inhomogeneous
broadening of the low temperature photoluminescence spectra (not shown
here). We have chosen the effective
masses of electrons and 
holes to be $m_{\mathrm{e}}=0.07m_{0}$ and $m_{\mathrm{h}}=0.3m_{0}$,
respectively, where $m_{0}$ is the free electron mass, which yields the
lowest optical transition corresponding to a
bright state excited in the transverse direction
($n_{\lon}=1,n_{\tra}=2$) at the energy $\Delta E=41$~meV above the
ground state. Fitting to the overall polarization value observed in
the experiment yields the heavy-light hole separation $\Delta
E_{\mathrm{lh}}=30$~meV, which is a reasonable value (actually,
however, this energy should be correlated with the dot height, which
we neglect). The results
correspond to the thermal distribution of electron and hole occupations
at 300~K, assuming weak enough excitation to keep the Fermi system
non-degenerate.  

While the overall DOP follows from the value of the
heavy-light hole energy splitting which is a fitting parameter, the
agreement of the features appearing in the energy dependence is
remarkable. Both in the theory and experiment, the
DOP is roughly constant in the area of the maximum
luminescence (820~meV in the experimental results). Slightly to the
higher energies, the DOP drops. This can be understood from Eq.~\eqref{pol}: At
higher energies, the states excited along the structure ($n_{\lon}>1$)
have a larger contribution, which reduces the polarization. Farther
towards the higher energies, the state excited perpendicular to the
structure ($n_{\tra}=2$) starts to contribute strongly. At 300~K,
the occupation of this state is non-negligible. According to
Eq.~\eqref{pol}, for low
values of $n_{\lon}$, the DOP of the luminescence from this state is
four times higher than that of the ground state. This leads to a clear
maximum in the DOP observed in the experiment and reproduced in the
model. On the low energy side of the luminescence maximum, the theory
predicts a 
decrease in the DOP due to the increasing contribution of larger dots,
while the ground state emission dominates. The discrepancy with
respect to the experimental value is due to the abrupt decrease of the
detection sensitivity below $0.78$~eV (InGaAs detector) causing the
increased uncertainty of the experimental results in the area of weak
luminescence signal.

In general, Eq.~\eqref{pol} yields many predictions for the DOP
dependence on various system parameters. For structures with a small
asymmetry ($D\sim L$), the DOP should increase with growing
$L$. However, this increase saturates and for strongly elongated
structures ($L\gg D$) at low temperatures and weak
excitations, where the ground state emission dominates, one finds
$P\sim1/D^{2}$, independent of $L$. The DOP should decrease and then
increase with both temperature and excitation power as both these
factors lead to an increased contribution from the excited states,
first along the structure (which decreases the DOP), then transverse
to the structure (which increases the DOP). Finally,
semi-speculatively and beyond the presented formal model, one can
expect that taller structures (larger $H$) should generally have
smaller heavy-light 
hole separations and therefore should show stronger polarization.

In conclusion, we have formulated a semi-quantitative model that
explains the polarization of light emitted by a nanostructure with a
strong in-plane asymmetry (a quantum dash) by relating it to the hole
subband mixing. The model yields a simple estimate of the DOP in terms
of the geometrical and material parameters. We 
have tested our model 
against a room temperature measurement performed on an ensemble of
InAs/InP QDashes. We have shown that the model not only accounts for the
overall DOP for a reasonable value of the assumed heavy-hole splitting
but also reproduces the features observed in the energy dependence of
the DOP and allows us to interpret them in terms of the contribution
to the luminescence from various excited states. Based on our model,
we have made further predictions for the dependence of the DOP on the
structure parameters and experimental conditions.

\setlength{\parindent}{0mm}
\textbf{Acknowledgements.}  The authors would like to thank the group
from the Department of Applied Physics, University of W{\"u}rzburg
(Alfred Forchel, Sven H{\"o}ffling and Sebastian Hein) for providing
the structures used in the experimental part of the study. The work
has been supported by the Polish MNiSW (Grant No. N N202 181238). 
A.~Musia{\l} acknowledges financial support within a fellowship
co-financed by the European Union within the European Social Fund.


\end{document}